\documentstyle[12pt,titlepage]{article}

\newcommand{\be}{\begin{equation}}
\newcommand{\ee}{\end{equation}}
\newcommand{\br}{\begin{eqnarray}}
\newcommand{\er}{\end{eqnarray}}

\newcommand{\half}{\frac{1}{2}}
\def\a{{\alpha}}

\def\D{{\Delta}}

\def\m{{\mu}}
\def\n{{\nu}}

\def\dirac#1{\setbox0=\hbox{$#1$}\rlap{\hbox to \wd0{$\hss\mkern1mu/\hss$}} 
\box0 }

\begin{document}

 \begin{titlepage}
\begin{center}

{\bf INFINITIES IN PHYSICS AND \\
TRANSFINITE NUMBERS IN MATHEMATICS}

P. Narayana Swamy ($\dagger$ )\\
Department of Physics, Southern Illinois University, Edwardsville, IL 
62026-1654\\
\end{center}
\begin{center}
{\bf Abstract}
\end{center}

Several examples are used to illustrate 
how we deal cavalierly with infinities and unphysical systems in physics. 
Upon examining these examples in the context of infinities 
from Cantor's theory of transfinite numbers, the only known mathematical 
theory of infinities, we conclude that apparent inconsistencies in physics are 
a result of unfamiliar and unusual rules obeyed by mathematical 
infinities. We show that a re-examination
of some familiar
limiting results in physics leads to surprising and unfamiliar conclusions.
It is not 
the purpose of this work to resolve the problem of infinities but the intent 
of this analysis is to point out that the study of real infinities in 
mathematics may be the first step towards delineating and understanding the 
problem of infinities in physics.

\noindent PACS numbers: 02.10.-v  $\; , \quad$   02.10.Jf $\;, \quad$  
02.90,+p $\; , \quad$ 03.50.De, \quad 05.90.+m\\
($\dagger$ ): electronic address: pswamy@siue.edu\\
\noindent Preprint SIUE/HEP-2/May  1999
 \end{titlepage}

Let us begin with the notion of a line charge in electrostatics in classical 
Maxwell theory. If a one dimensional line charge density $\lambda$ extends 
from $-\half L$ to $\half L$ along the $y-$axis in vacuum, then the electric 
field at a distance $z$ situated on the bisector of the line charge 
is [\ref{Griffiths}]  the familiar result
\be
{\bf E}(z) = \frac{1}{2 \pi \epsilon_0} \frac{\lambda L}{z \sqrt{z^2 + L^2}} 
\; \hat{{\bf z}},
\ee
where $q= \lambda L$ is the total charge distributed on the line. When $z >> 
L$, the field is
\be
{\bf E}(z) \approx \frac{1}{2 \pi \epsilon_0} \frac {q}{z^2 } \; \hat{{\bf z}}
\ee
and the line charge at a large distance behaves as if it were a point charge. 
When $L >> z$, i.e., the charge is  of infinite length, and the field reduces 
to
\be
{\bf E}(z) \approx \frac{1}{ 2\pi \epsilon_0} \frac{ \lambda }{z } \; \hat{{\bf z}}.
\ee
The standard expectation that the total charge on the line is infinitely 
large,
 the length is infinite so that the ratio $\lambda$ is a finite number cannot 
be proved by using ordinary mathematics of finite numbers. We may resolve this problem once and for all, by fiat, or by 
hypothesis, and define the density in this manner to be finite. This is the 
sort of assumption one makes in statistical mechanics on which all of the 
derivation of thermodynamics rests. It is a reasonable hypothesis,  and we shall 
return to it later.

The notion of a line charge is still an unphysical situation {\it e.g., \/} 
when we calculate the electrostatic energy. In classical electromagnetic theory, we 
calculate the electrostatic energy of a system by bringing in discrete charges 
one by one from a far away point (infinity). This method runs into difficulty 
because in the case of a line charge, the charge is already there at infinity. 
Moreover, in physically realistic systems, 
the field due to a charge configuration is expected to decrease as $1/r^2$ but the electric field due to an 
infinite line charge with uniform density $\lambda$ decreases as $1/r$ for 
large $r$: it is not fast enough to assure the existence, the 
finiteness that is,  of many physical quantities and hence the physical 
admissibility of 
an infinite line charge can be called into question. For an understanding of the  
full implications of this paradoxical situation, let us consider the fact 
that the electrostatic energy is given by the familiar expression
\be
W = \half \int \rho V d^3x \quad = \half \epsilon_0 \int (\nabla \cdot {\bf 
E}) V d^3x \quad = \half \epsilon_0 \int \left  [ \nabla \cdot (V {\bf E}) - 
{\bf E}\cdot (\nabla V)    \right ] d^3x
\ee
due to Poisson equation satisfied by ${\bf E}$. Employing Gauss' theorem, we 
arrive at the familiar result
\be
W = \half \epsilon_0 \int \;d^3x \;  {\bf E}\cdot {\bf E} ,
\label{5}
\ee
after these crucial steps: first extending the system volume to infinity and 
secondly  discarding the surface term which vanishes because for realistic 
systems, electric field decreases at least as fast as $1/r^2$ and the 
potential decreases at least as fast as $1/r$ while surface area increases no 
faster than $r^2$. In this context therefore, we find that, if we admit physical 
systems such as infinitely long line charges then the electric field does not 
decrease sufficiently fast for the surface integral term to vanish. For such 
systems the canonical expression, Eq.(\ref{5}) for the electrostatic energy density is then 
problematic. 
Moreover, a similar expression for the magnetostatic energy density will no 
longer be valid. This will lead to other not well-behaved physical quantities 
[\ref{rite}]
such as current density ${\bf J}$, then the vector potential ${\bf A}$ and so 
on and the entire edifice of classical electromagnetic theory is of dubious 
validity. 

Consequently 
we may concede that systems such as line charges are unphysical (in 
mathematical parlance, they do not exist). This amounts to an assumption 
that infinity does not exist and hence not a truly satisfactory 
resolution -- if we assume that there are real infinities, which is the major 
premise of this discussion. 

Let us turn our attention to the problem of determining magnetic fields of 
simple and standard electric current configurations by the 
application of Biot-Savart law. Due to a thin conductor of finite length $L$ 
carrying a current $I$, the magnitude of the magnetic field at a distance $r$ on the 
bisector is
\be
B = \frac{\m_0}{2\pi r} \frac{L}{\sqrt{L^2 + r^2}},
\ee
which reduces to the following for an infinitely long wire:
\be
B \approx  \frac{\m_0}{2\pi r}.
\ee
If we take the finite wire, cut in two and then evaluate the limit when $L >> 
r$ we obtain the same result as above. However, if we consider a 
semi-infinitely long wire, the magnetic field is only half of the above. 
This is a puzzling result if we take the length as a potential infinity. As a 
real infinity, it is equally puzzling because half of infinity appears to be 
different from infinity.

Let us consider another interesting result. If we take the finite wire and 
bend it around into a circle of radius $r$, then the magnitude of the magnetic field at the 
center of such a circular wire carrying a current $I$ is 
\be
B = \frac{\m_0 I}{2 r}
\ee
which would reduce to
\be
B = \frac{\m_0 I}{4 r}
\ee
if it is bent into a semicircle. The factor 2 or 4 is not significant here, 
neither the factor of $\pi$ (since the $\pi$ arising from circular functions 
occurring in the straight wire calculation gets cancelled by the different 
topology in the semicircle): what is really astonishing is that the result 
for a finite wire and an infinite wire are practically of the same form. What is going 
on here? We shall return to this problem later.

The position eigenfunction in quantum mechanics is a Dirac delta function [\ref{Sakurai}]  in 
coordinate space. This wave function is not square integrable, is indeed 
infinite. We deal with this problem by requiring that it be treated as a 
distribution rather than as an ordinary mathematical function.

Let us consider how we handle the problem of the electon self-energy which is 
infinite in classical Maxwell theory [\ref{Griffiths}]. It arises as a consequence of two facts: the electron is  a point 
particle, and the electric field is defined at every point in space. In 
classical physics it is a linear infinity and in quantum theory it is a 
logarithmic divergence. 
The resolution to the problem of infinite 
self-energy requires the full apparatus of renormalization in quantum 
electrodynamics. However, the essential argument can be stated in its most 
simple form as follows. The plane wave wavefunction representing an electron 
must include the self-energy and is thus
\be
\psi({\bf x},t)= e^{ (i/{\hbar}) [ {\bf p}\cdot {\bf x}- (E + \D E)t ]} .   
\ee
Observables which are bilinear in $\psi$ and 
$\psi^*$ are invariant under a  phase transformation. If we therefore perform 
the global
phase transformation
\be
\psi({\bf x},t) \; \rightarrow \; \psi' ({\bf x},t)= e^{ (i/{\hbar})   
\D E t },     
\ee
the transformed wavefunction is free of the self-energy and we have thus 
performed the equivalent of an unobservable canonical transformation. As a 
result of this transformation, the energy of the electron has become 
finite only because we have employed the hypothesis that infinity subtracted 
from infinity yields a finite quantity or zero. This result is not true of 
real infinities [\ref{upanishad}]. If we regard 
the self-energy as only a potential infinity, there are no rules in 
mathematical physics that we know of for such quantities and the rules for 
finite arithmetic are not reliable here. We may take  the standard 
mathematical rules of addition and subtraction established for finite 
quantities and employ them for infinite quantities with impunity. This 
approach will forever lead to contradictions and therefore the theory built on 
such manipulations will never be self-consistent. On the other hand, we may 
regard such infinities as real infinities, admit their existence. In the 
latter case, a self-consistent analysis is possible while recognizing  that 
infinities are somewhat peculiar and strange but we can make sense out of them 
if we understand that the usual rules of arithmetic do not apply to such 
quantities. We shall explore the latter point of view. 

Infinity also occurs in the theory of quantum harmonic oscillators. The energy 
spectrum  is discrete and the total energy is
\be
E = \sum_{\nu}(n + \half) h \n
\ee
where $n=0,1,2, \cdots$ and we must sum over all frequencies (discrete or continuum) from $0$ to 
$\infty$. The energy of the ground state corresponding to vacuum (state 
containing nothing)
diverges, which is the familiar zero-point energy of a quantum system. The 
customary procedure to deal with this infinity is to rescale the energy value. 
This is a reasonable procedure since in practice we measure only differences.
However, this procedure is again questionable since subtracting 
infinity from infinity does not necessarily lead to a finite result. Before 
proceeding further, let us discuss the case of statistical thermodynamics.

It is remarkable that thermodynamics deals with a very large number of 
microscopic constitutents (we might regard them as infinitely many, in the 
potential sense) and yet predicts finite results which thus make physical 
sense. In the application of statistical mechanics to physical systems where 
the goal is to derive the thermodynamic properties, we follow an important 
procedure: we evaluate everything in the thermodynamic limit. That is, we 
allow the number of particles $N$ to go to infinity, and allow the physical 
volume $V$ of the system to go to infinity but define the ratio, namely the 
density $N/V$ to be always finite. Many physical quantities such as 
temperature, entropy all remain finite. This subtle concept is especially made 
poignant in several phenomena. One interesting example is the case of Bose 
condensation. The occupation number of a Bose-Einstein system consisting of 
many atoms of mass $M$ is given by
\be
N = \frac{1}{e^{\a} -1} \; + \; \frac{V (2M)^{3/2}} {2 \pi^2 \hbar^3} \; 
\int_0^{\infty}\; \frac{\sqrt{E}}  { e^{  \a + E/kT}     -1}.
\ee
where $\a = - \m /kT$. Isolating the ground state in this manner is essential 
because otherwise the ground state would be accompanied by zero probability 
because of the density factor $\sqrt{E}$ which in turn would have disallowed 
condensation by assumption. Everything is fine for $\a >0$ but when $\a=0$, 
the first term will dominate over the second term and the ground state will 
have macroscopically large probability resulting in condensation. The result 
for the occupation number  $N_e$ of the excited states is given by
\be
N_e = N \left \{ 1 -  \left( \frac{T}{T_c } \right )^{3/2}  \right \}.
\ee
This result is arrived at only [\ref{Lawrie}]  in the thermodynamic   limit $N 
\rightarrow \infty, \; V \rightarrow \infty$   and shows that $N_e$ becomes 
very small as the temperature decreases to the critical temperature and $N_0 = 
N - N_e$ becomes marcroscopically large. It is remarkable that, unlike quantum 
mechanics or electromagnetic field theory, Statistical thermodynamics is a 
clean theory not bedeviled by problems arising from infinities or unphysical 
concepts. The remarkable thing about systems described by statistical 
thermodynamics is this: $N$ and $V$ are large in the sense of the macroscopic in 
contrast to microscopic scale, and one has to employ mathematical limits to 
implement this criterion, but they are not truly infinite. Indeed they are 
finite as can be calculated from the parameters namely, mass density, atomic 
weight and Avogadro number.  Consequently the problem of infinities simply 
does not arise in thermodynamics and solid state physics. 

The study of infinities in mathematics is the first step towards 
understanding the problem of infinities in physics. There has not been a great 
deal of attention devoted to this fundamental mathematical problem probably 
because mathematicians do not believe in real infinities and physicists just 
wish it away {\it i.e.,\/}  mathematicians 
believe that infinities do not exist, with one exception: Georg Cantor 
[\ref{Cantor}] who is credited with inventing set theory, 
developed the theory of transfinite numbers. We shall now present a brief 
review of Cantor's theory of infinities.

Cantor's study of transfinite numbers was inspired by Galileo's paradox 
[\ref{Pi}]: the 
set of natural numbers $1, 2, 3, \cdots , N, \cdots $ can be placed  in one-to-one 
correspondence with the squares of natural numbers $1, 4, 9, \cdots , N^2, 
\cdots$, each set being endless. The numbers in the second infinite set will 
appear somewhere in the first set and thus the second infinite set is a subset 
of the first infinite set. Rather than concluding that there is a 
contradiction, inspection shows that the rules for  infinite numbers are 
different from those of finite numbers. Cantor thus defines infinity as a 
collection of objects which can be put in one-to-one correspondence with a 
part of itself. Cantor then proceeds to introduce the notion of cardinality, a 
measure of how many actual numbers belong in the set. The cardinality of 
natural numbers is defined as $\aleph_0$. Considering the set of natural 
numbers $1, 2, 3, \cdots $ and the set formed by adding zero to  even numbers, 
one immediately establishes a one-to-one correspondence. Thus we have Cantor's 
first theorem:
\be
\aleph_0 + 1 = \aleph_0,
\ee
which, by similar argument, can be extended to
\be
\aleph_0 + f = \aleph_0
\ee
where $f$ is any finite number. This leads to the conclusion that infinity 
subtracted from infinity is not necessarily finite but could be indeterminate. 
This argument can be extended to an arbitraily large number $f$. Indeed it is 
true that infinity subtracted from infinity may also remain infinite, as can 
be established by the famous story [\ref{Pi}]  of Hilbert's hotel.

The second important theorem due to Cantor deals with the question of the 
cardinality of fractions. Although it appears as if there exist infinitely 
many fractions between two natural numbers, a careful counting [\ref{Pi}, 
\ref{Hersh}] establishes the 
result that the set of numbers including fractions is of the same cardinality 
as the set of natural numbers. Between each of the natural numbers we have 
$\aleph_0$ set of fractions and consequently, the theorem:
\be
\aleph_0^2 = \aleph_0, \;  \aleph_0^3= \aleph_0, \cdots , \;  \aleph_0^r=\aleph_0, 
\cdots .
\ee
Cantor's third important theorem address the question of the cardinality of real 
numbers [\ref{Cantor}, \ref{Pi}]. Cantor's analysis establishes the result that the reals have a higher 
cardinality. Denoting the cardinality of reals which includes rational as well 
as irrational numbers by $C$ for continnum [\ref{aside}], we have 
Cantor's theorem:
\be
C + \aleph_0= C.
\ee

Cantor has shown how to develop higher cardinalities [\ref{Cantor}]. Cantor's 
fourth 
theorem involves infinities in more than one dimension. One 
would think at the outset that sets in higher dimensions would be of higher 
cardinality than $\aleph_0$ but one can prove that they have the same 
cardinality and thus the theorem:
\be
C \times C = C, \qquad C \times C \times C = C, \;   \cdots,
\ee
a theorem so incredible that even Cantor could not believe what it signifies.

At this point, we can ask what all this implies for physics and physical 
systems. There are significant implications for physical systems if we accept 
the major premise that rather than being merely potential, real infinities do 
exist. We shall now cite some examples to illustrate some of these 
implications.

The case of an infinitely long line charge distribution is a simple example. 
In this case both the length and the electric charge are of cardinality $C$ 
for the continuum and consequently the ratio $\lambda$ representing the charge 
density is finite. The case of the thermodynamic limit invoked in statistical 
mechanics is a fascinating counter-example. If we treat them as real 
infinities, then the ratio $N/V$ involves the 
number of atoms which is countable and thus of cardinality $\aleph_0$ whereas 
the volume of the system is of cardinality $C \times C \times C= C$. For this 
ratio to be finite we must require that $C $ must equal a finite number times 
$\aleph_0$. This last statement is however, not true for real infinities. 
Hence it follows that the finiteness of the number density $N/V$ would not 
have been possible if $N$ and $V$ were real infinities. Indeed the finiteness 
of the number density follows from the fact that each of these quantities is 
finite to begin with, as was discussed earlier.

Consider the case of the infinitely long straight wire carrying an electric 
current. Clearly, the points on this wire can be put in one-to-one 
correspondence with the points on a finite semicircular wire by drawing the 
radii emanating from the center of the semicircle reaching out to the wire. In 
this manner we find that the cardinality of the finite semicircular wire and 
the infinitely long wire are the same, namely $C$, thus mapping $- \half L, 
\half L$ on to $0, \pi$. It is thus not at all surprising that the  magnetic field 
due to these is of the same form: $B= \m_0 I/ 2 \pi r$ compared to $B=\m_0 I/ 
4R$ (we already discussed the origin of the factor of $\pi$ ). What is a 
puzzle according to finite mathematics is thus seen to be a direct consequence 
of Cantor's transfinite mathematics.

There is an interesting implication of Cantor's theorem 3 for the problem of 
the quantum system of a particle in a box. The wavefunction describing the 
one-dimensional problem with the length of the box   $L$ is
\be
\psi(x) = \sum_1^{\infty} A_n \sin \frac{n \pi x}{L}.
\ee
The modes are labeled by the integer quantum number $n = 1, 2, 3, \cdots , 
\infty$. If the number of modes indeed reach infinity (real not just 
potential), then its cardinality is $\aleph_0$. Note that the coordinate $x$ 
is of cardinality $C$ but $n$ is of cardinality $\aleph_0$. The states or 
modes of the 
system are described by the quantum number $n$. The wavefunction for a 
corresponding three dimensional system of dimensions $L_1, L_2, L_3$ is
\be
\psi(x,y,z)= \sum A_{{\ell}, m, n} \sin \frac{ {\ell}\pi x}{L_1} \; \sin 
\frac{m \pi y}{L_2} \; \sin \frac{n\pi z}{L_3},
\ee
where the modes are described by the endless list of independent integer 
numbers $\ell, m, n = 1, 2, \cdots, \infty$. While the multiplicity for finite 
values is three-fold compared to the one-dimensional problem, the cardinality 
of the infinite system is no more than $\aleph_0$ due to Cantor's fourth 
theorem. Extending this to the study of the hydrogen atom energy levels 
described by Bohr theory in one dimension and Sommerfeld theory in three 
dimensions we observe the following. In the Bohr theory, the energy levels are 
enumerated by the principal quantum number $n = 1, 2, 3, \cdots \infty$ with 
the infinity signifying the continnum. In the Sommerfeld theory, the energy 
levels are enumerated by the three quantum numbers, $n,\; \ell = 0, 1, 2, \cdots 
n-1$ and $m_{\ell} =- \ell, -\ell + 1, \cdots  0, 1, 2, \cdots \ell$. The 
multiplicity associated with the principal quantum number $n$ turns out to be 
$\sum (2 \ell +1) = n^2$. (Note that only $n$ is given by an endless list). In 
the case $n \rightarrow \infty$, (potential) the usual expectation is that the 
number of sublevels is of a higher order infinity. However according to 
Cantor's fourth theorem, they both are of the same cardinality and hence in 
the case $n = \infty$ we obtain the puzzling result: $n^2 =  n$.

In the analysis of many problems in classical electromagnetic theory, we come 
across the standard argument where the volume integral of a divergence of a 
string of quantities involving a vector is converted to a surface integral of 
the normal component which is then discarded in the limit when the system 
volume is taken as infinity, thus
\be
\int_V \; d^3x \; \frac{\partial}{\partial x_i} (A_i B_j C_k D_l T_{jkl} \cdots)\; 
= \; \int_S\; d^2x
\; n_i S_i \; \approx 0.
\ee
This result is true only for the infinitely large 
volume of the system which implies that we are dealing with a real infinity 
otherwise we cannot perform these manipulations self-consistently. If that is 
the case we may observe by Cantor's theorem 4  that the cardinality of the 
surface integral is the same as the volume, namely the continuum set $C$. 
Hence one is not anymore negligible than the other and consequently the 
standard results derived by employing finite mathematics are suspect. 

The problem of self-energy is much more complex. Since the infinity arises 
here because of the point charge as well as the fact that fields are defined 
at every space point, the interpretation of the infinity is more difficult, 
and the resolution is also difficult, Cantor's transfinite mathematics 
notwithstanding.

We conclude by making a few general observations. 

Thermodynamic systems are devoid of infinities and are inherently finite. $N$ 
is countable and large 
and $V$ is measurable and macroscopically large but all physical parameters are finite and 
measurable and finite, including the number density. There is a class of 
physical systems containing infinities but which can be re-examined by using 
methods which have successfully prevailed in thermodynamics and statistical 
mechanics, with a view to resolve the problem infinities in these systems. 

There is a class of physical phenomena where infinities occur which may 
require revolutionary ideas in order to deal with the infinities. Well-known 
examples are a) infinite intensity predicted by classical theory (ultraviolet 
catastrophe) in cavity radiation which was resolved by Max Planck in 1900 b) 
infinite density predicted by classical theory of the hydrogen atom due to 
radiative collapse resolved by Niels Bohr in 1915. Systems containing point 
sources and fields at every space-time point apparently belong to this class. 
It can also be argued that fields are open systems which are inherently 
infinite and thus infinities are natural ingredients in these systems.

Finally there may be physical systems containing real infinities which cannot 
be transformed away. These systems may perhaps be understood only by a 
re-examination based on Cantor's transfinite mathematics. In this context, it 
is useful to remember that Cantor's theory is well grounded in physical 
reality: it is based on arithmetic and set theory. This may help us understand 
the apparent contradictions and problems of self-consistency. At any rate, 
unerstanding infinities as mathematical entities is the first step in,  and a 
prerequisite to, understanding infinities in the real physical world.

{\bf REFERENCES AND FOOTNOTES}
\begin{enumerate}
\item \label {Griffiths} See e.g., D. Griffiths, {\it Introduction to 
Electrodynamics\/}, Prentice-Hall, (1999) Upper Saddle River NJ.

\item \label{rite} Considering their unphysical nature, it is paradoxical that 
the problems such as the infinitely long line charge, dielectric sphere in an 
omnipresent uniform field and others are all rites of passage in the standard 
undergraduate physics curriculum. 

\item \label{Sakurai} J. J. Sakurai, {\it Modern Quantum Mechanics\/}, 
Benjamin Cummings Publishing Co. (1985) Reading, MA.

\item \label{upanishad} S. Radhakrishnan, {\it Principal Upanishads\/} Chapter 
5, page 289 (1968), G. Allen and Unwin Ltd. London. A sanskrit poem provides a 
succinct definition of infinity and the relevant part can be roughly 
translated as: "If infinity is taken away from the infinite, what remains is 
still infinite".

\item \label{Lawrie} Ian Lawrie, {\it A Unified Grand Tour of Theoretical 
Physics\/}, Adam Hilger (1990), Bristol.

\item \label{Cantor} Georg Cantor, {\it Contributions to the founding of the 
theory of transfinite numbers\/}, translated by P. Jourdain, (1915), Dover 
Publications Inc. New York NY.

\item \label{Pi} J. Barrow, {\it Pi in the Sky\/}, Clarendon Press, Oxford 
(1992).

\item \label{Hersh} Reuben Hersh, {\it What is Mathematics, Really ?\/}, 
Oxford University Press (1997) New York NY.

\item \label{aside} An interesting point in the history of mathematics is the 
continuum hypothesis: that there is no cardinality bigger than $\aleph_0$ and 
smaller than $C$. This notion is neither proved nor disproved. See {\it 
e.g.,\/} reference \ref{Hersh}.

\end{enumerate}

\end{document}